\documentclass[prl,twocolumn,amsmath,amssymb,nofootinbib]{revtex4-1}
\usepackage[T1]{fontenc}
\usepackage{amsthm}
\usepackage{graphicx}
\usepackage{times}
\usepackage{hyperref}
\hypersetup{
breaklinks=true,
colorlinks=true,
linkcolor=red,
citecolor=red,}
\usepackage{textcomp}
\usepackage{xcolor}
\usepackage{enumerate}
\usepackage{multirow}
\usepackage{tabularx}
\usepackage[mathscr]{euscript}
\usepackage{mathtools}
\usepackage[most]{tcolorbox}
\usepackage{needspace}
\usepackage[mathscr]{euscript}
\usepackage{dsfont}
\usepackage{amsfonts}
\usepackage[normalem]{ulem}
\usepackage{centernot}
\usepackage{stmaryrd}
\usepackage{longtable}
\usepackage{tikz}
\usetikzlibrary{quantikz}

\newcommand\dmu[1]{\int{d #1}}
\newcommand\ddmu[2]{\int{d #1 d #2}}
\newcommand\s[1]{_{\rm #1}}
\newcommand\sn[2]{_{{\rm #1}{;}#2}}
\newcommand\us[1]{^{(#1)}}

\renewcommand{\bra}[1] {\langle #1 |}
\renewcommand{\ket}[1] {| #1 \rangle}
\newcommand{\ketbra}[1]{ | #1 \rangle\!\langle #1 |}

\newcommand{\Tr} {\operatorname{Tr}}
\newcommand{\poly} {\operatorname{poly}}

\newcommand{\one}{\leavevmode\hbox{\small1\normalsize\kern-.33em1}}

\newcommand{\CC}{{\cal C}}

\newcommand{\G}{{\cal G}}

\newcommand{\MM}{\mathsf{M}}
\newcommand{\pp}{\mathsf{P}}
\newcommand{\TT}{\mathsf{T}}
\newcommand{\cc}{\mathsf{C}}

\newcommand{\prob}{\mathfrak{P}}
\newcommand{\alg}{\mathfrak{A}}

\begin{document}
\title{Quantum computational advantage implies contextuality}
\author{Farid Shahandeh}
\email{Electronic address: shahandeh.f@gmail.com}
\affiliation{Department of Physics, Swansea University, Singleton Park, Swansea SA2 8PP, United Kingdom}

\begin{abstract}
We show that a separation between the class of all problems that can efficiently be solved on a quantum computer and those solvable using probabilistic classical algorithms in polynomial time implies the generalized contextuality of quantum algorithms.
Our result subsumes versions of Gottesman-Knill theorem as special cases.
\end{abstract}

\maketitle

Quantum computations promise significant speedups over classical ones~\cite{Preskill2012,Deutsch1992,Shor1994,Aaronson2013}.
This embarks on the challenge of understanding the physical property responsible for such advantages.
The answer turns out to be elusive, leaving the problem largely open.
On the one hand, there are several models of quantum computations, e.g., the circuit model~\cite{NielsenBook} and the measurement-based quantum computation~\cite{Raussendorf2001,Raussendorf2003,Nielsen2003,Leung2004}.
On the other hand, many properties such as coherence~\cite{Streltsov2017}, entanglement~\cite{Horodecki2009}, nonlocality~\cite{EPR,Bell1964,Bell2004}, and contextuality~\cite{KS,Bell1966,Spekkens2005} are genuinely quantum, the relationships between which are not fully understood.
Thus, depending on one's choice of the quantum computational model, one or some of these quantum properties may appear fundamental for quantum computational advantage~\cite{Matera2016,Jozsa2003,Vidal2003,Gross2009,Bravyi2018,Raussendorf2003,Howard2014,Bermejo-Vega2017}.

Among the possibilities, quantum contextuality~\cite{KS,Spekkens2005} has recently drawn interest as an appealing resolution~\cite{Raussendorf2003,Howard2014,Bermejo-Vega2017}.
The advantage of contextuality for computations was noted by Spekkens~\textit{et al.}~\cite{Spekkens2009} in the limited scenario of parity-oblivious multiplexing.
Later, it was shown that measurement-based-quantum computations~\cite{Raussendorf2001,Anders2009} capable of computing nonlinear Boolean functions with the assistance of a classical linear computer provide proofs of contextuality~\cite{Raussendorf2013,Frembs2018}.
The necessity of contextuality in universal quantum computations, however, was demonstrated in the restricted model of quantum computation by state injection~\cite{Howard2014,Delfosse2015,Bermejo-Vega2017,Raussendorf2017,Karanjai2018,Schmid2021}.

In this paper we will make the notion of generalized (non)contextuality of quantum circuits precise.
We show that efficiently solving a classically hard problem on a quantum processor certifies generalized contextuality~\cite{Spekkens2005} of the quantum algorithm.
Therefore, genaralized contextuality is the necessary resource for any quantum computational speedup.
Specifically, we provide a classical \textit{weak}-simulation algorithm that efficiently samples from the output distribution of a quantum circuit that admits a generalized noncontextual hidden-variable model (here and after, called \textit{ontological} model) on an instance of size $n$.
If the quantum algorithm admits a noncontextual ontological model (NCOM), then our algorithm suffices to exclude any superpolynomial quantum speedup.
We find that the noncontextuality assumption is restrictive enough to abandon quantum computational advantages regardless of the ontic spaces' cardinality.
Interestingly, our result highlights that there could exist nonseparable ontological models for quantum algorithms that are classically efficiently simulatable due to being noncontextual.
Hence, quantum contextuality is a stronger requirement for obtaining quantum computational advantage than entanglement.


\paragraph*{Circuit Model.---}
Consider the circuit model of quantum computations~\cite{Jozsa2003,NielsenBook}.
The computational unit is defined to be a qubit that can be prepared in either of the computational basis states $\ket{0}$ or $\ket{1}$.
In each instance of the computation, the integer $n$ denotes the size of the input, i.e. the number of qubits used.
Operations on qubits are performed through a finite gate set $\G{=}\{U_{i}\}_{i=1}^{K}$ consisting of $K$ unitary transformations, each of which acts on at most two qubits.
A circuit family $\CC{=}\{C_n\}$ is such that each of its elements is indexed by the size $n$ of its input and is constructed by the serial and parallel compositions of the gates in $\G$.
The family is called poly-sized if the number of gates making up each circuit $C_n$ is upper-bounded by a polynomial in $n$ and uniform if a Turing machine can output its classical description in a time that is upper-bounded by a polynomial in $n$.
The complexity class of all computational problems which can be solved using such circuits with a bounded error probability is called {\rm BQP}~\cite{NielsenBook}.  

Given an instance $\mathbf{i}{=}i\us{1}\cdots i\us{n}{\in}\{0,1\}^n$ of size $n$, the computation proceeds as follows.
The $n$ qubit input is prepared in a product state $\ket{\mathbf{i}_0}{=}{\bigotimes}_{i=1}^n\ket{i_0\us{i}}$.
Then the circuit $C_n$ is applied to the input quantum state in a sequence of $D(n){=}\poly (n)$ time steps where in each step at most $n$ gates are applied in parallel.
The computation is finished by the measurement of $m{\leq} n$ qubits in the computational basis and tracing out the rest.
The probability of an outcome $\mathbf{x}{=}x_1\cdots x_m{\in}\{0,1\}^m$ in the measurement $\mathbf{M}(m)$ which corresponds to the projection onto $\bra{\mathbf{x}}{=}{\bigotimes}_{i=1}^m\bra{x\us{i}}$ and encodes the result of the computation is given by
\begin{equation}\label{eq:qcomp}
\begin{split}
& p(\mathbf{x}|\mathbf{i}_0,C_n,\mathbf{M}) = \\
& \Tr \ketbra{\mathbf{x}} U_{D;k_D}\cdots U_{D;1}\circ \cdots \circ U_{1;k_1}\cdots U_{1;1}[\mathbf{i}_0],
\end{split}
\end{equation}
Here, $\mathbf{i}_0$ is short for the projector $\ketbra{\mathbf{i}_0}$ and $U_{s;k_s}$ is understood as the $k_s$th gate applied in the $s$th time step of the circuit.
Also, the symbol $\circ$ represents concatenation of gates in different time steps.


\paragraph*{Ontological Models.---}
Underlying the notion of noncontextuality are hidden-variable or, more specifically, ontological models~\cite{Spekkens2005}.
They are founded on the assumption that the information specifying preparations of physical systems do, in fact, determine an underlying {\it ontic} state $\lambda$ which corresponds to the real states of affairs of the system. 
The transformations set out how $\lambda$ is manipulated upon acting on the system, and measurement procedures indicate which outcome will be obtained due to the system's ontic state.
Let us denote by $\pp$, $\TT$, and $\MM$ the information uniquely specifying these elements, respectively, and by $\Lambda$ the space of all ontic states.
Any uncertainty in the precise state of the system, whether due to fundamental epistemic restrictions or ignorance, is represented by an uncertainty in the exact value of $\lambda$.
Hence, preparations, transformations, and measurements are most generally identified by probability distributions $\mu(\lambda|\pp)$, stochastic maps $\tau(\lambda'|\lambda,\TT)$, and indicator functions $\xi(x|\lambda,\MM)$ for each outcome $x$, respectively.
These functions are nonnegative for all $\lambda$, $\lambda'$, and $x$, and satisfy $\dmu \lambda\, \mu(\lambda|\pp){=} 1$, $\dmu\lambda\,\tau(\lambda'|\lambda,\TT){=}1$, and $\sum_{x}\xi(x|\lambda,\MM){=}1$ for all $\lambda$.
These properties ensure the probabilistic nature of the model.
The probability of the outcome $x$ in a measurement $\MM$ given the preparation $\pp$ and transformation $\TT$ is obtained according to the rule 
\begin{equation}\label{eq:Bayes}
p(x|\pp,\TT,\MM) = \ddmu\lambda\lambda' \xi(x|\lambda',\MM)\tau(\lambda'|\lambda,\TT)\mu(\lambda|\pp).
\end{equation}
It is implicit that ontological models are $\lambda$-sufficient, i.e., the system’s relevant properties which determine the output distribution of a transformation or the outcome of a measurement are captured entirely by the ontic state $\lambda$.
As such, $\lambda$ screens off any of the preparations, transformations, and measurements from the other two~\cite{Harrigan2010}.


\paragraph*{Noncontextuality of Quantum Circuits.---}

The \textit{broad noncontextuality}~\cite{Shahandeh2021} is a hypothesis motivated by Leibniz's methodological principle of \textit{the identity of empirical indiscernibles}~\cite{Spekkens2019}, stating that whenever two or more specifications of an operation give rise to indiscernible statistics, they must be represented identically and uniquely in any mathematical model of the experiment.
Quantum theory agrees with this hypothesis.
For instance, a horizontal polarization state $\ketbra{H}$ of a photon does not encode any information regarding its preparation procedure, e.g., the photon being passed through a polarizing filter or reflected off of a dielectric surface.
Nor does it include any redundant parameters irrelevant to the observable statistics of the system, e.g., the temperature of a far away galaxy.

Quantum circuits are physical processes that can primitively be described via operational elements, i.e., the preparations, transformations, and measurements~\cite{Spekkens2005}.
In terms of these elements, one can represent an \textit{instance} of a quantum circuit as
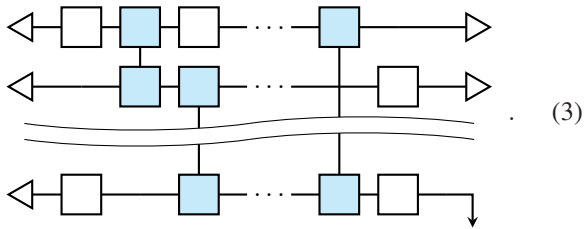
\begin{equation}\label{eq:QCS}
\begin{quantikz}[row sep=0.4cm]
\prepT{} & \gate{} & \gate[style={fill=cyan!20}]{}\vqw{1} & \gate{} 		   & \ \ldots\ \qw & \gate[style={fill=cyan!20}]{} \vqw{3} & \qw     & \meterT{} \\
\prepT{} & \qw     & \gate[style={fill=cyan!20}]{}        & \gate[style={fill=cyan!20}]{} 		   & \ \ldots\ \qw & \qw             & \gate{} & \meterT{} \\
\wave & & & & & & & \\
\prepT{} & \gate{} & \qw            & \gate[style={fill=cyan!20}]{} \vqw{-2} & \ \ldots\ \qw & \gate[style={fill=cyan!20}]{}         & \gate{} & \trash{}
\end{quantikz}.
\end{equation}
Here,$
\begin{quantikz}[yshift=0.08cm]
\prepT{}
\end{quantikz}$
denotes a preparation---there are $n$ independent preparations.
Each single box denotes a single qubit transformation and each pair of connected boxes represents a two-qubit transformation.
$
\begin{quantikz}[yshift=0.08cm]
\meterT{}
\end{quantikz}$
denotes a measurement on a qubit and there are $m{\leq} n$ of them.
The downward arrow represents unmeasured qubits.
Importantly, the parameter dependencies among elements of Eq.~\eqref{eq:QCS}, when exist, are specified by the mathematical model (quantum theory or the ontological model) describing the circuit.

Suppose that an ontological model underlies the circuit of Eq.~\eqref{eq:QCS} and that $\cc_n$ is a specific operational recipe for implementing $C_n$.
We impose the broad noncontextuality hypothesis on the preparations of the instance by requiring that for two distinct preparation procedures $\pp_1$ and $\pp_2$, $p(\mathbf{x}|\pp_1,\cc_n,\MM) {=} p(\mathbf{x}|\pp_2,\cc_n,\MM)$ for all $\mathbf{x}$, $\cc_n$, and $\MM$ if and only if $\mu(\lambda|\pp_1){=}\mu(\lambda|\pp_2)$.
When satisfied, we say the circuit is preparation noncontextual.
This definition is inspired by that of Spekkens~\cite{Spekkens2005} and justified by the intuition that two distinct preparations of the same instance do not alter the output statistics because they yield the same probability distribution at the ontic level.

Now suppose that $\cc_1$ refers to a specific recipe for implementing a specific part of $\cc_n$ and denote the remaining of the circuit by $\overline{\cc}_n$.
Following the intuition above, we say the circuit is transformation noncontextual, i.e., $p(\mathbf{x}|\pp,\cc_1,\overline{\cc}_n,\MM) {=} p(\mathbf{x}|\pp,\cc_2,\overline{\cc}_n,\MM)$ for all $x$, $\pp$, $\overline{\cc}_n$, and $\MM$, if and only if $\tau(\lambda'|\lambda,\cc_1){=}\tau(\lambda'|\lambda,\cc_2)$.
Here, $\cc_2$ is a different recipe for implementing the same part of the circuit as $\cc_1$.

Finally, we say the circuit is measurement noncontextual, i.e., for two distinct measurement procedures $\MM_1$ and $\MM_2$, $p(\mathbf{x}|\pp,\cc_n,\MM_1) {=} p(\mathbf{x}|\pp,\cc_n,\MM_2)$ for all $\mathbf{x}$, $\pp$, and $\cc_n$, if and only if $\xi(\mathbf{x}|\lambda,\MM_1){=}\xi(\mathbf{x}|\lambda,\MM_2)$.

\paragraph*{\textbf{Definition~1.}}
An instance of a quantum circuit is ontologically noncontextual if and only if its operational description as in Eq.~\eqref{eq:QCS} admits an NCOM, i.e., an ontological model which is preparation, transformation, and measurement noncontextual.

\paragraph*{\textbf{Definition~2.}}
A quantum computation algorithm $\alg$ is ontologically noncontextual if and only if there exists a constant $N$ such that every instance of every circuit of size $n {>} N$ admits an NCOM.

Recall that quantum theory is ontologically contextual~\cite{KS,Spekkens2008,Ferrie2008}.
We emphasize, however, that this is not a good reason for believing that quantum circuits for all quantum algorithms are ontologically contextual.
In fact, there are many counterexamples, e.g., the optical circuits that allow for nonnegative phase-space representations~\cite{Bartlett2002,Rahimi-Keshari2018}.

Quantum computations promise that they are strictly more powerful than classical computations.
More precisely, it is believed that there exists at least a problem $\prob$ that admits a polynomial-time (in the input size $n$) quantum algorithm that solves all the instances of size $n$ (for all $n$), while it lacks a polynomial-time classical algorithm.
Formally, $\prob{\in}{\rm BQP}$ but $\prob{\notin}{\rm BPP}$, where ${\rm BQP}$ and ${\rm BPP}$ are the bounded-error quantum polynomial time and bounded-error probabilistic polynomial time complexity classes.
This assumption refutes the extended Church-Turing thesis and admits a separation between quantum and best classical algorithms for solving particular problems.

The main result of this paper is the following theorem that connects the quantum-classical computational separation to the fundamental hypothesis of noncontextuality.

\paragraph*{\textbf{Theorem.} }
A separation between $\rm BQP$ and $\rm BPP$ implies the ontological contextuality of the quantum computation algorithm $\alg$ that efficiently solves ${\rm BQP}\ni\prob\notin{\rm BPP}$.

Note that, we do not assume that the procedure implementing the algorithm $\alg$ is a universal quantum computation, as a BosonSampler is not~\cite{Aaronson2013}.

Our proof is by contradiction and can be sketched as follows: noncontextuality of transformations and their invertibility (when restricted to each circuit), imply that they must be deterministic.
This, together with the noncontextuality assumption, imposes a restriction on the structure of the ontic space underlying each circuit.
Then, the separability of the input quantum state and the readout measurements together with the structure required for the ontic space and noncontextuality, imply that input ontic distribution and the indicator function at the output must factorize.
These properties enable an efficient classical simulation algorithm for ontologically noncontextual circuits, refuting their ability to provide any quantum speedup.
We now provide the details of the proof.

%
\paragraph*{Proof of Theorem.---}

Central to our proof is the following lemma.

\textit{\textbf{Lemma.}}
Every gate of an ontologically noncontextual circuit must be represented by a deterministic transformation.

It is true that in every given circuit $\cc_n$ every gate $\TT$ is invertible with respect to the circuit.
This means that, there is a $\TT^{-1}$ such that inserting the concatenations $\TT{\circ}\TT^{-1}$ or $\TT^{-1}{\circ}\TT$ at any point of the circuit does not change the output statistics, hence the result of the computation.
We note that this is weaker than $\TT$ being invertible with respect to the full quantum theory.
This, together with the noncontextuality of the transformations in an NCOM, imply that the ontological transformation $\tau(\lambda'|\lambda,\TT)$ is invertible for each gate so that $\tau(\lambda'|\lambda,\TT^{-1}) {=} \tau^{-1}(\lambda'|\lambda,\TT)$, at least when restricted to the ontological representation of the instance of the circuit.
It is, however, trivial that for every restrictively invertible map $\tau {:} \{\mu\} {\to} \{\mu\}$ there is an extension $\tau {:} \Lambda {\to} \Lambda$ to a stochastic map that is invertible over the whole ontic space.
Finally, we know that invertible stochastic maps are deterministic in both finite and infinite dimensions~\cite{Farahat1966,Schwarz1967,Attal2010} which completes the proof of the lemma.
There is also a physically intuitive explanation for the latter statement, namely that, invertible stochastic maps preserve the entropy~\cite{Mackey1992} implying their deterministic nature.

Now suppose that there exists an efficient quantum algorithm $\mathfrak{A}$ for the problem $\prob{\in}\rm BQP$ with $\prob {\notin} \rm BPP$. 
Suppose also that an NCOM is known for every instance of the circuits given by $\alg$ for all 
$n$ greater than some constant $N$.
We show that this would imply an efficient classical simulation of the quantum circuits corresponding to the algorithm.

We begin with analysing the properties of the ontic space $\Lambda$ that underlies an arbitrary noncontextual quantum circuit $C_n$.
Consider any single qubit gate $U^{(i)}$ in the circuit that acts on $i$th qubit.
Using the lemma above, we know that, such unitary is represented by a deterministic ontic transformation $\pi^{(i)}(\lambda'|\lambda,U^{(i)})$ in the noncontextual ontological model.
Let us denote the invariant subspace on which the group of such single qubit transformations acts nontrivially by $\Lambda^{(i)}$. 
Now, the noncontextuality assumption dictates that for any pair $i$ and $j$ of qubits, their respective nontrivial invariant subspaces $\Lambda^{(i)}$ and $\Lambda^{(j)}$ must be disjoint.
It thus follows that the complete space must be of the form $\Lambda {=} \Lambda^{(i)} {\times} \cdots {\times} \Lambda^{(n)}\times\Lambda\s{nl}$.
The subspace $\Lambda\s{nl}$ denotes the space of holistic properties of the $n$-qubit system and contains all the information that is not available to individual qubits locally.

Similar to single-qubit gates, each two-qubit gate $U^{(ij)}$ between qubits $i$ and $j$ is also represented by a deterministic transformation $\pi^{(ij)}(\lambda'|\lambda,U^{(ij)})$ within the noncontextual ontological model.
Denote the invariant \textit{two-qubit} subspace on which the group of such transformations acts nontrivially by $\Lambda^{(ij)}$.
Given the local spaces of qubits $i$ and $j$ we find that this subspace can most generally be of the form $\Lambda^{(ij)}{=}\Lambda^{(i)}{\times}\Lambda^{(j)}{\times}\Lambda\s{nl}^{(ij)}$ where $\Lambda\s{nl}^{(ij)}$ encodes the holistic two-qubit properties and may get populated only if a joint operation is performed on qubits $i$ and $j$.
Appealing to the noncontextuality assumption once more, the size of the local two-qubit ontic space $\Lambda^{(ij)}$ is independent of the input size, so is the size of $\Lambda\s{nl}^{(ij)}$.
It then follows automatically that for two distinct pairs of qubits $k,l {\neq} i,j$ the subspaces $\Lambda^{(ij)}$ and $\Lambda^{(kl)}$ are disjoint.

Repeating the above argument for all single qubits and all pairs of qubits, we find that $\Lambda {=} \Lambda\s{QCS}{\times}\overline{\Lambda}\s{nl}$ where $\Lambda\s{QCS} {=} \prod_i\Lambda^{(i)}\prod_{i<j}\Lambda\s{nl}^{(ij)}$.
Here, $\overline{\Lambda}\s{nl}$ represents the nonlocal subspace of $\Lambda$ that is accessible only if a joint operation is performed on more than two qubits.
The latter subspace is evidently irrelevant to our purpose because we are restricted to at most two-qubit operations at a time.
We thus see that, while ontological noncontextuality does not imply separability of the ontic space, it imposes a strong restriction on its structure.

We now describe the algorithm that simulates the computation if an NCOM is given.
The quantum input $\ket{\mathbf{i}_0(n)}{=}\ket{i_0\us{1}}{\otimes}{\cdots}{\otimes}\ket{i_0\us{n}}$ to the circuit is specified ontologically by $\mu_0(\lambda|\mathbf{i}_0) {=} \mu_0\us{1}(\lambda^{(1)}|i_0\us{1}) {\times} \cdots {\times} \mu_0^{(n)}(\lambda^{(n)}|i_0\us{n})$ where $\lambda^{(i)}{\in}\Lambda^{(i)}$.
This can be deduced from the structure of $\Lambda$, the noncontextuality assumption stating that ontic representations cannot depend on statistically irrelevant parameters, and that the input preparations are independent:
Consider two qubits $i$ and $j$.
First, the ontic distribution of qubit $i$ cannot depend on the choice of preparation of qubit $j$, i.e., $\mu_0\us{i}(\lambda^{(i)}|\lambda^{(j)},\lambda^{(ij)}\s{nl},i_0\us{i},i_0\us{j})=\mu_0\us{i}(\lambda^{(i)}|\lambda^{(j)},\lambda^{(ij)}\s{nl},i_0\us{i})$, and vice versa.
This is a form of \textit{parameter independence}~\cite{Emerson2013} and follows from the noncontextuality of preparations~\cite{SpekkensTalk_pirsa2012}.
Second, the ontic distribution of qubit $i$ cannot depend on the ontic state of qubit $j$, i.e., $\mu_0\us{i}(\lambda^{(i)}|\lambda^{(j)},\lambda^{(ij)}\s{nl},i_0\us{i}) {=} \mu_0\us{i}(\lambda^{(i)}|\lambda^{(ij)}\s{nl},i_0\us{i})$, and vice versa.
The latter resembles the \textit{outcome independence}~\cite{Emerson2013}.
Third, the ontic distribution of qubit $i$ cannot depend on the holistic properties because there has been no joint operation on qubits $i$ and $j$ yet.
Hence, $\mu_0\us{i}(\lambda^{(i)}|\lambda^{(ij)}\s{nl},i_0\us{i})=\mu_0\us{i}(\lambda^{(i)}|i_0\us{i})$, and similarly for qubit $j$.

It is important to note, however, that noncontextuality does not defy the possibility of entangled states during the computation because the nonlocal subspaces $\Lambda^{(ij)}\s{nl}$ may get populated as a result of two-qubit gates.

Similarly, the ontological representation of the projector $\ketbra{\mathbf{x}} = \ketbra{x\us{1}} \otimes \cdots \otimes \ketbra{x\us{m}} \otimes I^{\otimes n-m}$ at the end of the circuit is specified by $\xi(\mathbf{x}|\lambda,\mathbf{M})=\xi^{(1)}(x\us{1}|\lambda^{(1)},M\us{1})\times\cdots\times \xi^{(m)}(x\us{m}|\lambda^{(m)},M\us{m})\prod_{i=m+1}^{n} \big[\xi^{(i)}(0|\lambda^{(i)},M\us{i}) + \xi^{(i)}(1|\lambda^{(i)},M\us{i})\big]$. 
By the same reasoning, a single-qubit gate $U_s^{(i)}$ on qubit $i$ and a two-qubit gate $U_s^{(ij)}$ between qubits $i$ and $j$ at the time step $s$ are given by deterministic maps $\pi_s^{(i)}(\lambda'^{(i)}|\lambda^{(i)},U_s^{(i)})$ and $\pi_s^{(ij)}(\lambda'^{(ij)}|\lambda^{(ij)},U_s^{(ij)})$, respectively, where $\lambda^{(i)},\lambda'^{(i)}\in\Lambda^{(i)}$, and $\lambda^{(ij)},\lambda'^{(ij)}\in\Lambda^{(ij)}$.

The simulation algorithm then proceeds as follows.
\paragraph*{1.~a.}
Sample $\lambda_0 {=} (\lambda_0^{(1)},\dots,\lambda_0^{(n)}) {\in} \prod_{i=1}^n\Lambda^{(i)}$ according to the probability distribution $\mu_0(\lambda|\mathbf{i}_0)$ by sampling from each subsystem independently.
\paragraph*{1.~b.}
Extend $\lambda_0$ to $\overline{\lambda}_0 {\in} \Lambda\s{QCS}$ by appending zeros to it for nonlocal subspaces.
\paragraph*{2.~a.}
At step $s {=} 1$, for each application of the single-qubit gate $U_1^{(i)}$ on $i$th qubit, update the sample point according to the deterministic transformation $\pi_1^{(i)}(\lambda'^{(i)}|\lambda^{(i)},U_1^{(i)})$ to $\overline{\lambda}_1^{(i)}$ which is simply obtained by the coordinate substitution $\lambda_0^{(i)}\mapsto\lambda'^{(i)}_1$.
\paragraph*{2.~b.}
For each two-qubit gate $U_1^{(jk)}$ between qubits $j$ and $k$ ($j,k {\neq} i$) at step $s{=}1$, update the point $\overline{\lambda}_1^{(i)}$ to $\overline{\lambda}_1^{(ijk)}$ by the substitution 
$\lambda_0^{(jk)}=(\lambda_0^{(j)},\lambda_0^{(k)},\lambda\sn{nl}{0}^{(jk)})\mapsto (\lambda_1'^{(j)},\lambda_1'^{(k)},\lambda\sn{nl}{1}'^{(jk)})=\lambda'^{(jk)}_1$ 
according to the transformation $\pi_1^{(jk)}(\lambda'^{(jk)}|\lambda^{(jk)},U_1^{(jk)})$.
\paragraph*{2.~c.}
Repeat the processes \textit{2.~a} and \textit{2.~b} for all gates in $s{=}1$ to obtain the complete updated point at this step, denoted by $\overline{\lambda}_1$.
\paragraph*{3.}
Repeat the same processes \textit{2.~a} through \textit{2.~c} for steps $1{<}s{\leqslant} D$ to get the output ontic state $\overline{\lambda}_D$.
\paragraph*{4.}
Return the output $\mathbf{x}$ according to the distribution $\xi(\mathbf{x}|\lambda_D,\mathbf{M})$ in which $\lambda_D$ is the projection of $\overline{\lambda}_D$ onto the locally available ontic subspace $\prod_{i=1}^n\Lambda^{(i)}$.

By appealing to the promise that the noncontextual ontological model reproduces exactly the predictions of quantum mechanics via the relation Eq.~\eqref{eq:Bayes}, one can readily verify that the probability of each sample outcome $\mathbf{x}$ using this classical algorithm equals that of quantum algorithm in Eq.~\eqref{eq:qcomp}.

The simulation algorithm above is a classical stochastic one and is efficient.
To see this, we note that it is possible to sample the distribution in step \textit{1.~a.} classically efficiently.
This is guaranteed by the fact that the ontic input distribution is factorizable and the assumption that we can sample efficiently from the specified initial single-qubit ontic distributions. 
Since the ontic transformations are known and the cardinalities of single- and two-qubit ontic subspaces are independent of $n$, implying the number of parameters required to keep track of the ontic (sample) state $\overline{\lambda}$ scales only as $O(n^2)$, the state update steps of the algorithm are efficient.
Finally, the indicator function of the measurements is factorizable and known that implies the efficiency of the sampling in step \textit{4}.

Notably, the parameter space needed to characterize the entire ontic space $\Lambda$ may still be exponentially large due to the nonlocal part $\overline{\Lambda}\s{nl}$ that encodes holistic properties of three or more qubits.
This subspace, however, does not populate because we are only restricted to single- and two-qubit operations and assume that the model is noncontextual.

Indeed, the presumable noncontextual ontological model for the computational algorithm can be assumed to be a continuous variable one.
In that case, the algorithm above assumes infinite precision real numbers that may only run on rather unrealistic analog classical computers~\cite{Vergis1986}.
Nevertheless, it is possible to discretize the model which enables an approximate simulation algorithm.
Notably, in one such algorithm the deterministic transformations are essentially translated into permutations.
Then, it is a standard practice to show that bounded-error input distributions, stochastic transformations, and readouts lead to an efficient classical sampling of the output probability with bounded error~\cite{Veitch2012,Mari2012}.
We thus conclude that our classical simulation algorithm for an arbitrary circuit is efficient if an ontologically noncontextual model can be provided.
This contradicts the assumption that there is a separation between BPP and BQP, hence the result.
\qed


\paragraph*{Discussion and Conclusions.---}

We have shown that generalized contextuality is a necessary resource for quantum speedups in the circuit model of quantum computations.
This is true even if onological models with arbitrary large ontic spaces are adopted.
Our proof rests on the observation that if a quantum algorithm admits an NCOM, then there is an efficient classical algorithm that samples from its output distribution to arbitrary precision.

From a different perspective, the separation between the quantum and classical computational powers has physical consequences, the same way violating Bell inequalities for classical correlations has implications for our fundamental understanding of Univers’s working principles.
We found that, if the conjecture is true, any superpolynomial quantum speedup certifies the violation of the noncontextuality assumption.

Many no-speedup results in the literature are special cases of our theorem.
For example, in the Gottesman-Knill theorem for continuous-variable systems~\cite{Bartlett2002,Bartlett2003}, the transformations are symplectic ones~\cite{Simon1994} that act deterministically on the Weyl-Wigner phase space points.
The resulting bosonic quasiprobability representations~\cite{GlauberBook} is a natural NCOM for ciruits
built from symplectic transformations and restricted to input states and measurements with nonnegative quasiprobability representations, implying their efficient classical simulatability.
A similar example can be given for discrete-variable systems of odd-prime dimensions~\cite{GrossThesis,Howard2014}. 

Our result also sheds light on the role of entanglement in quantum speedups.
The Gottesman-Knill theorem shows that even highly entangling circuits may be efficiently classically simulatable.
By virtue of our analysis, the presence of entanglement in circuits may only correspond to the population of two-qubit nonlocal ontic subspaces.
Such subspaces may be sufficient for the construction of NCOMs allowing for efficient classical simulation algorithms.
This is in agreement with the findings of Jozsa and Linden~\cite{Jozsa2003} that bounded multipartite entanglement renders quantum circuits classically efficiently simulatable.

Finally, Lillystone \textit{et al.}~\cite{Lillystone2019} have recently shown that the classically simulatable single-qubit stabilizer theory is generalized contextual.
Then, they have argued that such prevalence and weakness of generalized contextuality makes it unfavourable to be a candidate resource for quantum speedups.
However, we believe that a relevant parameter to be analyzed for drawing such conclusions is the scaling of quantum resources with the circuit size rather than the mere presence or absence of the resource.
Our results paves the way towards such analyses.


\begin{acknowledgements}
The author gratefully acknowledges Richard Jozsa, Martti Karvonen, and Roohollah Ghobadi for valuable discussions and comments.
This project was supported by the Royal Commission for the Exhibition of 1851 Research Fellowship. 
\end{acknowledgements}

\bibliography{ContextQC}
\bibliographystyle{apsrev4-1new}

\end{document}